\documentclass{elsart}
\usepackage{natbib}
\usepackage{graphicx}
\usepackage{amsmath,amssymb,amsfonts,appendix}
\begin{document}

\begin{frontmatter}

\title{
How to grow a bubble:\\
A model of myopic adapting agents}

\author{Georges Harras \& Didier Sornette}

\begin{center}
{Chair of Entrepreneurial Risks\\ Department of
Management, Technology and Economics \\ ETH Zurich, CH-8001 Zurich,
Switzerland}
\end{center}

\begin{abstract}

We present a simple agent-based model to study the development of a bubble and the consequential
 crash and investigate how their proximate triggering factor
 might relate to their fundamental mechanism, and vice versa. Our agents invest according to
 their opinion on future price movements, which is
 based on  three sources of information, (i) public information, i.e. news, (ii)
 information from their ``friendship'' network and (iii) private information.
Our bounded rational agents continuously adapt their
 trading strategy to the current market regime by weighting each of these sources
 of information in their trading decision according to its recent predicting performance.
We find that bubbles originate from a
 random lucky streak of positive news, which, due to a feedback mechanism
 of these news on the agents' strategies develop into a transient collective herding regime.
 After this self-amplified exuberance, the price has reached an unsustainable high value,
 being corrected by a crash, which brings the price even below its fundamental value.
 These ingredients provide a simple
 mechanism for the excess volatility documented in financial markets. Paradoxically, it is
 the attempt for investors to adapt to the current market regime
 which leads to a dramatic amplification of the price volatility.
 A positive feedback loop is created by the two dominating mechanisms
 (adaptation and imitation) which, by reinforcing each other, result in bubbles and
 crashes. The model offers a simple reconciliation of the two opposite  (herding versus
 fundamental) proposals for the origin of crashes within a single framework and justifies
 the existence of two populations in the distribution of returns, exemplifying the concept
 that crashes are qualitatively different from the rest of the price moves.

\end{abstract}


\end{frontmatter}

\section{Introduction}
\label{sec:intro}

Bubbles and crashes in financial markets are events that are fascinating to academics and
practitioners alike. According to the consecrated academic
 view that markets are efficient, bubbles, being temporally persistent, self-reenforcement deviations
 of the price from the fundamental value, are impossible. And crashes should only result from
 the revelation of a dramatic piece of information.
 Yet in reality, there is a large consensus both from professionals \citep{Dudley2010,Trichet2010} and academia
 \citep{shiller_2000,Abreu2003Bubbles} that bubbles do exist,
 and even the most thorough post-mortem analyses are typically inconclusive as to
 what piece of information might have triggered the observed crash \citep{krach87}.

It is often observed that crashes occur soon after a long run-up of prices, referred to as a bubble.
A crash is thus often the burst of the bubble. There is a vast amount of literature aiming at
 characterizing the underlying origin(s) and mechanism(s) of financial bubbles
 \citep{Abreu2003Bubbles, kaufman_2001, sheffrin_2005, shiller_2000, sornette_2003} but there is still
 no consensus in the academic community on what is really a bubble and what are its
 characteristic properties. Bubbles do not seem to be fully explained by bounded rationality
 \citep{levine07}, speculation \citep{Lei2001Nonspeculative} or the uncertainty in the
 market \citep{Smith1988Bubbles}. Finally, there is no really satisfactory
 theory of bubbles, which both encompasses its different possible mechanisms
 and adheres to reasonable economic principles (no arbitrage, equilibrium, bounded rationality, ...).

Most approaches to explain crashes search for possible mechanism or effects
 that operate at very short time scales (hours, days, or weeks
 at most). Other mechanisms concentrate on learning an exogenously given crash rate
 \citep{Sandroni1998Learning}. Here, we build on the radically different hypothesis summarized in
 \citep{sornette_2003} that the underlying cause of the crash should be found in the
 preceding months and years, in the progressively increasing build-up of a
 characteristic that we refer to as `market cooperation', which expresses the growth
 of the correlation between investors' decisions leading to stronger
 effective interactions between them as a result of several positive feedback mechanisms.
 According to this point of view, the proximal triggering factor for price collapse
 should be clearly distinguished from the fundamental factor. A crash occurs
 because the market has entered an unstable phase toward the culmination of
 a bubble and any small disturbance or process may reveal
 the existence of the instability. Think of a ruler held up vertically on your finger: this
 very unstable position will lead eventually to its collapse, as a result of a small (or an absence
 of adequate) motion of your hand or due to any tiny whiff of air. This is the proximal cause
 of the collapse. But the fundamental cause should be attributed to
the intrinsically unstable position.

What is then the origin of the maturing instability?
Many studies have suggested that bubbles result from the over-optimistic expectation
 of future earnings and history provides a significant number of examples of bubbles driven by such unrealistic
 expectations \citep{Kindleberger2005Manias,sheffrin_2005,sornette_2003}. These
 studies and many others show that bubbles are initially nucleated at times of burgeoning
 economic fundamentals in so-called ``new economy'' climates. This vocable refers
 to new opportunities and/or new technological innovations. But, because there are large
 uncertainties concerning present values of the economies that will result from the
 present innovations, investors are more prone to influences from their peers  \citep{ThyNeighborsPortfolio2005}, the media,
 and other channels that combine to build a self-reflexive climate of (over-)optimism \citep{Umpleby07}.
 In particular, these interactions may lead to significant imitation, herding and collective
 behaviors. Herding due to technical as well as behavioral mechanisms
 creates positive feedback mechanisms, which lead to self-organized
 cooperation and the development of possible instabilities or to the ``building
 of castles in the air'', to paraphrase Malkiel \citep{malkiel_1990}.
 This idea is probably best exemplified in the context of the Internet bubble
 culminating in 2000 or the recent the CDO bubble in the USA peaking in 2007, where the new
 economies where the Internet or complex derivatives on sub-prime mortgages building
 on accelerating real-estate valuations.

Based on these ideas, the present paper adds to the literature by providing a detailed analysis of how the proximate triggering factor of
a crash might relate to its fundamental mechanism in terms of a global cooperative herding mechanism.
 In particular, we rationalize the finding of \cite{cutler-1989} that exogenous news are responsible for no more than
 a third of the variance of the returns and that major financial crises are not preceded by any particular dramatic news.

In a nutshell, our multi-period many agent-based model is designed as follows.
At each time step $t$, each investor forms an opinion
on the next-period value of a single stock traded on the market. This opinion is shaped by
 weighting and combining three sources of information available at time $t$: (i) public
 information, i.e. news, (ii) information from their ``friendship'' network, promoting
 imitation and (iii) private information. In addition, we assume that the agents adapt
 their strategy, i.e., the relative importance of these different sources of information
 according to how well they performed in the past in predicting the next-time
 step valuation.

The a priori sensible qualities of our agents to gather all possible information and adapt
to the recent past turn out to backfire. As their decisions are aggregated in the market, their collective impact
leads to the nucleation of transient phases of herding with positive feedbacks. These nucleations occur as a result of
random occurrences of short runs of same signed news.
Our main findings can thus be summarized as follows: rallies and crashes occur due
to random lucky or unlucky streaks of news that
are amplified by the feedback of the news on the agents' strategies into collective transient herding regimes.
In addition to providing a convincing mechanism for bubbles and crashes, our model also
provides a simple explanation for the excess volatility puzzle \citep{Shiller1981Do}.

Before presenting the model and its results, it is useful to compare it with the relevant literature and related models.
A related line of research aims at developing a theory of ``convention''
 \citep{Orlean1,Orlean2,Orlean3,Orlean4,Orlean5,Orlean6}, which emphasizes that even the
 concept of ``fundamental value'' may be a convention established by positive and
 negative feedbacks in a social system. A first notable implementation by
 \cite{Topol1991Bubbles} proposes a model with an additive learning process between
 an `agent-efficient' price dynamics and a mimetic contagion dynamics. Similar to our own set-up, the agents of  \cite{Topol1991Bubbles}
 adjust their bid-ask prices by combining the information from the other buyers' bid prices, the other sellers' ask prices and the agent's
 own efficient price corresponding to his knowledge of the economic fundamentals.  \cite{Topol1991Bubbles} shows that mimetic
 contagion provides a mechanism for excess volatility.
 Another implementation of the concept of convention by  \cite{Wyart2007Selfreferential} shows that
 agents who use strategies based on the past correlations between some news and returns may actually produce
 by their trading decisions the very correlation that they postulated, even when there is no a priori economic basis for such correlation.
 The fact that agents trade on the basis of how the information forecasts the return is reminiscent of our model, with however
 several important differences. The first important conceptual change is that \cite{Wyart2007Selfreferential}
 use a representative agent approach (in contrast with our heterogeneous agent framework), so that
 effect of imitation through the social network is neglected. The second difference is in
 the agent's calculation of the correlation to adapt their strategies. In
 \citep{Wyart2007Selfreferential}, agents' strategies are controlled by the correlation between
 the news and the return resulting immediately from their aggregate action based on those news (taking into account the agents'
 own impact). Our agents' strategies are determined by the correlation between their information and the return one time step later,
 which embodies the more realistic situation, in which a postion first has to be open and then
 closed a time step later for the trade's payoff to be observed.

Another closely related line of research is known as ``information cascades''.
According to  \citep{Bikhchandani1992Theory},
 ``an informational cascade occurs when it is optimal for an individual having observed the
 action of those ahead of him, to follow the behavior of the preceding individual without
 regard to his own information''. In these models, agents
 know that they have only limited information and use their neighbors actions in order to
 complement their information set.  \cite{Bikhchandani1992Theory} showed that the fact that agents
 use the decisions of other agents to make their own decision will lead with probability
 $1$ to an informational cascade under conditions where
 the decisions are sequential and irreversible. This model was
 later generalized by \cite{Orlean6} into a non-sequential version, where
 informational cascades were still found to be possible.

The concept of information cascades is not new in modeling bubbles.
 \cite{Chari2003Hot} developed a model where agents try to compensate their uncertainty
 about the a priori fixed payoff of an asset by observing all other agents' actions.
 In our model, agents are also using the opinions of their neighbors to
 determine how to act but the reason behind this is different.
 Our agents are not so much interested in the fundamental value of the stock, but more
 in its future directions. They try to buy the asset before its price rises and sell before it
 falls, making profit from the difference in the price.
The true underlying equilibrium
 value is not the only important information to them, and they are more clever than purely
 fundamental value investors. They recognize that fundamental value is just
 one component among others that will set the market price. They include the
 possibility that the price may deviate from fundamental value, due to other behavioral
 factors. And they try to learn and adapt to determine what are the dominant factors.
 In principle, they should be able to discover the fundamental value and
 converge to its equilibrium. But it is a fact that they do not in some circumstances,
 due to the amplification of runs of positive or negative news in the presence of their
 collective behavior when sufficiently strong.
In the ``information cascade'' set-up,
 one assume that the ``truth'' exists, that there is a true
 fundamental price or a correct choice to be made which is exogenously given,
 and agents have no influence on the outcome. In our model however the outcome,
 whether selling or buying a stock was the right choice, is endogenously emerging from the
 aggregated choices of all agents. There is no a priori right or wrong answer, it is decided during the process.
 Moreover, the strength of the influence of her neighbors onto a given agent is not constant in time.
 This influence by the social environment evolves in time according to its past relevance and success.

A model for the formation of a boom followed by a crash was also developed by
 \cite{Veldkamp2005Slow}, where the price of an unknown company can rise only slowly
 due to infrequent news coverage. If the company performs well resulting in a slow
 boom, its susceptibility towards news increases as the media become more
 aware of the successful company so that, eventually, a single piece of bad news can induce
 a sudden crash. Although the subject of research is the same, we show how a boom can also be
 formed with news not being constantly positive and that a single piece of bad news does not
 necessarily lead to the burst a bubble.

The endogenization of the sources of information onto the decisions of the agents
 is inspired by the model of  \cite{Zhou2005Selffulfilling}, which
 focuses on herding and on the role of  ``irrational'' mis-attribution of price moves
 to generate most of the stylized facts observed in financial time series. Similarly to their model as well
 as many other artificial financial market models investigating the interaction between trading agents,
 our model is based on the Ising model,
 one of the simplest models describing the competition
 between the ordering force of imitation or contagion and the disordering impact of private
 information or idiosyncratic noise that promotes heterogeneous decisions \citep{Ising1}.

Our paper is organized into four sections. In Section 2, the detailed working of the
 model is presented. The results are shown and discussed in section 3 and section 4 concludes.

\section{The model}
\label{sec:model}

\subsection{General set-up}\label{sec:general_set-up}

We consider a fixed universe of $N$ agents who are trading (buying or selling)
 a single asset, which can be seen as a stock, the market portfolio or any other
 exchange traded asset. This asset is traded on an
 organized market, coordinated by a
 market maker. At each time step, agents have the possibility to either trade or to remain passive.
 The trading decision of a given agent is based on her opinion on the future price development.

To form their opinion, agents use information from three
 different sources: idiosyncratic opinion, global news and their network of acquaintances.
 In order to adapt their decision making process to the current market situation,
 they are weighting the different information sources
 by their respective past predicting performance.
Limited to these sources of information, our agents act rationally, i.e., they use all information
available to them to maximize their profits. Since they use backward looking adapting strategies
with finite time horizons, our agents are
boundedly rational, with limited competence, resources and available time.

A limitation of the model is to assume that agents do not have access to more liquidity
than their initial wealth and that generated by their investments. Moreover, our universe
has a fixed population, so that there is no flux of new ``foreign'' investors that may be attracted
in the later stage of a bubble, and who could inflate it up
further \citep{sornette_zhou_2004,zhou_2006, zhou_2008}. We thus purposefully remove
one of the mechanisms, namely the increasing credit availability and credit creation \citep{caginalp_2001}, which has often
been reported as an important ingredient to inflate historical bubbles
\citep{Galbraith,sornette_2003_chap_3,Kindleberger2005Manias}. This allows us to focus
on the role of decision processes with conflicting pieces of information in the presence of local adaptation.

\subsection{Three sources of information}

At every time step, agents form anticipations concerning the future price movements based on three
 sources information.

A first source of information of a given agent is her private information, $\epsilon_i(t)$, which may reflect the
unique access to information not available publicly or the idiosyncratic, subjective view of the
particular agent on how the stock will perform in the future.
 The private information
 is different for every agent, is taken uncorrelated across agents and time: the innovations
$\epsilon_i(t)$ are normally distributed ($\epsilon_i(t) \sim N(0,1))$ and  i.i.d.

A second source is the public information, $n(t)$. Public information includes economic,
 financial and geopolitical news that may influence the future economic performance of the stock.
 To capture the idea that the public news, $n(t)$, is fully informational with no redundancy  \citep{chaitin_1987},
 we take $n(t)$ as a white Gaussian noise with unit
 variance, uncorrelated with the private information $\{\epsilon_i(t), i=1, ..., N\}$ of the agents.
 Although news are generated as a stationary process, we will see that their impact
on the agents evolves because of the adaptive nature of the agents' strategies.

The third source of information for a given agent is provided by the expected decisions of other agents
to whom she is connected in her social and professional network. With limited
 access to information and finite computing power (bounded rationality), it can be shown to be
 optimal to imitate others \citep{Orlean2,roehner_2000}. Moreover, there is clear empirical evidence that
 practitioners do imitate their colleagues  \citep{ThyNeighborsPortfolio2005}. In our model, agents gather information on the
 opinions of their neighbors in their social network and incorporate it as an ingredient into
 their trading decision.

Incorporating agent interaction in the opinion formation process leads to dynamics
 described by models derived from the Ising model. Many earlier works have already borrowed concepts from
 the theory of the Ising models and of phase
 transitions to model social interactions and organization,
 e.g. \citep{Follmer1974RandomEconomies,PhysicsToday,Montroll}. In particular, Orl\'ean \citep{Orlean1, Orlean2,
 Orlean3, Orlean4, Orlean5, Orlean6} has captured the paradox of combining rational and imitative
 behavior under the name ``mimetic rationality,'' by developing models of mimetic contagion of
 investors in the stock markets which are based on irreversible processes of opinion forming.

\subsection{Opinion formation}

Using the three sources of information described in the previous section, the opinion of agent $i$
at time $t$, $\omega_i(t)$, consists of their weighted sum,
\begin{equation}
\omega_i(t)=c_{1i}\sum^J_{j=1}k_{ij}(t-1)\, E_i[s_j(t)]\,+\,c_{2i}\, u(t-1)\,
n(t) \, + \, c_{3i}\, \epsilon_i(t),
\label{eq:opinion-form}
\end{equation}
where $\epsilon_i(t)$ represents the private information of agent $i$, $n(t)$ is the public
information, $J$ is the number of neighbors that agent $i$ polls for their opinion and $E_i[s_j(t)]$
is the expected action of the neighbor $j$ estimated by agent $i$ at time
$t$\footnote{We use a
sequential updating mechanism with a random ordering. In this way, when agent $i$
polls her neighbors, she has a mix of opinions coming from those who have already
updated theirs and those have not yet. This procedure can be thought of as a device
to account for the large distribution of reactions times of humans
\citep{Vasquez_et_al_06,craneschsor10}.}.
The functional form of expression (\ref{eq:opinion-form}) embodies our hypothesis that an agent
 forms her opinion based on a combination of different sources of information. This is a standard
 assumption in the social interaction literature \citep{CambridgeJournals:463236, Brock2001Discrete}
 and decision making theory (see for instance \citep{Kording07}).

To take into account the heterogeneity in trading style and preferences of traders, we assume
that each agent $i$ is characterized by a triplet of fixed traits, in the form of the
 weights  $(c_{1i}, c_{2i}, c_{3i})$ she attributes to each of the three pieces of information
(social network, news and idiosyncratic).
The values $(c_{1i}, c_{2i}, c_{3i})$ for each agent are chosen randomly
from three uniform distributions over the respective intervals
 $[0,C_1]$, $[0,C_2]$ and $[0,C_3]$, at the initialization of the system. In section
 \ref{sec:trading_decision}, we will extend this heterogeneity by allowing for different
 risk aversions.

In order to adapt to the recent market regime, each agent can modify the weights she attributes to
the information from each of her neighbor $j$, via the factor $k_{ij}(t)$, and to
the public news, via the factor $u(t)$. The factors $k_{ij}(t)$'s and $u(t)$
 are updated such as to give more weight to an information source if it was a good predictor in the
 recent past, and to decrease its influence in the inverse case (more details in sec.~\ref{sec:adaptation}).
 The idiosyncratic term is not weighted and has a constant impact on agents actions.

Finally, for simplicity, our agents live on a virtual square lattice with $J=4$ neighbors,
 with periodic boundary conditions. The reported results are not sensitive to this topology,
 and hold for random as well as complete graphs.

\subsection{Trading decision} \label{sec:trading_decision}

Until now, we have introduced heterogeneity between agents through their three
personal traits ($c_{1i}$, $c_{2i}$ for $c_{3i}$), unique to each agent, on how they combine
information to form their opinion. Another important well-documented heterogeneity is
that different people have different risk aversions. We capture this trait by assuming that
each agent is characterized by a fixed threshold $\underline{\omega_i}$, controlling
the triggering of an investment action, given her opinion level $\omega_i(t)$.
An agent $i$ decides to go long (buy a stock) if
her conviction $\omega_i(t)$ is sufficiently positive so as to reach the threshold:
$\omega_i(t) \geq \underline{\omega_i}$. Reversely,
she decides to go short (sell a stock) if $\omega_i(t) \leq -\underline{\omega_i}$.
Thus, we assume symmetric levels of conviction in order for a trade to occur either on
the buy or sell sides. The parameter
 $\underline{\omega_i}$ captures one dimension of the agent's risk aversion: how much
 certitude she needs to break her hesitation and move into the market. The larger her
threshold $\underline{\omega_i}$, the larger certitude about future price
 movements the agent requires in order to start trading. Each agent is characterized
 by a different $\underline{\omega_i}$, drawn randomly from a uniform distribution in the interval
 $[0,\underline{\Omega}]$.

 As previously discussed in section \ref{sec:general_set-up},
our agents are liquidity constrained. The portfolio of an agent $i$ at time $t$ is the sum of
her cash $cash_i(t)$ and of the
 number $stocks_i(t)$ of the single asset that is traded in our artificial market.
 When an agent decides to buy, she uses a fixed fraction $g$
 of her cash. When an agent decides to sell, she sells the same fixed fraction $g$ of the value of her stocks.
 The fact that $g$ is much smaller than $1$ ensures time diversification. Our main
 results do not change significantly as long as $g$ does not approach $1$.
 Our agents are not allowed to borrow, because they can only buy a new stock, when
 they have the cash. Reciprocally, we impose short-sell constraints, in the sense that
 an agent can only sell a stock she owns. Thus, our model is related to the literature
 investigating the role of short-sale constraints \citep{miller_1977, chen_2002, ofek_2003}.

These rules can be summarized in terms of the
direction $s_i(t)$ of the trading decision and the volume $v_i(t)$ (in units of
 number of stock shares) of the agent $i$:
\newline
\begin{tabular}{lrcl}
 - if $\omega_i(t)\,>\,\,\,\underline{\omega_i}$&: $\textrm{ }s_i(t)$&=&$+1$~(buying)\\
&$v_i(t)$&=&$g \cdot \frac{cash_i(t)}{p(t-1)}$\\
 - if $\omega_i(t)\,<-\underline{\omega_i}$&: $\textrm{ }s_i(t)$&=&$-1$~(selling)\\
&$v_i(t)$&=&$g \cdot stocks_i(t)$,\\
\end{tabular}
\newline
\newline
where $p(t)$ is the price of the asset at time $t$.
 When an agent is buying assets, her order volume $v_i(t)$ is determined by her available
 cash and by the stock share price $p(t-1)$ at the previous time step
 (the main results remain unchanged if agents would use the expected $p(t)$ instead).
Our agents are submitting market orders, such that
 the price to pay to realize an order is the
 new price $p(t)$ determined by the market maker. This new price
 is determined by the price clearing mechanism that aggregates the excess demand after all the
 traders have submitted theirs decisions.

\subsection{Price clearing condition}
\label{sec:price_clearing_condition}

Once all the agents have decided on their orders, the new price of the asset is determined
 by the following equations:
\begin{eqnarray}
r(t)&=&\frac{1}{\lambda\cdot N} \sum_{i=1}^Ns_i(t)\cdot v_i(t) \label{eq:return-calc}
\label{eq:price-clearing1}\\
\log\left[ \textrm{price($t$)}\right] &=&\log\left[\textrm{price($t-1$)}\right]+r(t)~,
\label{eq:price-clearing2}
\end{eqnarray}
where $\lambda$ represents the relative impact of the excess demand upon the price, i.e. the market
 depth. Similar to \citep{Beja1980Dynamic, Wyart2007Selfreferential},
 we neglect all higher order contributions in expression (\ref{eq:price-clearing1}) and
 use a linear market impact function, as a rough
 approximation at time scales significantly larger
 than the tick-per-tick time scales for which nonlinear impact functions are observed
 \citep{Plerou2002Quantifying}.

Expressions (\ref{eq:price-clearing1}) and (\ref{eq:price-clearing2}) can be interpreted in two ways. One
 is that the trading is performed through a market maker, disposing of an unlimited amount of cash
 and stocks. Agents submit all their market orders to the market maker, who, after adapting the
 price to the excess demand, executes all the agents' trades. Because the market maker adapts the
 price before he executes the trades, he has a competitive advantage and
 gets on average a significant positive return for his service.

An alternative interpretation is that the trading style of our agents is midterm to longterm
 trading, excluding high-frequency traders like hedge-funds and such. Once our agents have absorbed
 their information and taken a trading decision, the price has already changed due to faster
 agents using similar trading information.

\subsection{Cash and stock positions}

We assume a frictionless market with no transaction fees. Once the return and the new price are
determined by the market clearing equations
(\ref{eq:price-clearing1},\ref{eq:price-clearing2}), the cash and number of stocks held by each agent $i$
are updated according to
\begin{eqnarray}
cash_i(t)&=&cash_i(t-1)-s_i(t)\,v_i(t)\, p(t) \label{eq:cash-update}\\
stocks_i(t)&=&stocks_i(t-1)+s_i(t)\,v_i(t).\label{eq:stock-update}
\end{eqnarray}

\subsection{Adaptation} \label{sec:adaptation}

As described above, agents have pre-existing heterogeneous beliefs on the reliability
of the three different sources of information, quantified by their
three traits $c_{1/2/3i}$. In addition, we assume that agents adapt their belief
concerning the credibility of the news $n(t)$ and their trust in the advice $E_i[s_j(t)]$
of their social contacts, according to time-dependent weights $u(t)$ and
 $k_{ij}(t)$, which take into account their recent past performance. Specifically,
 an agent estimates the value of a source of information by the
 correlation between the source's prediction and the realized return.
 For their strategy to be adapted to the current market regime, agents
 prioritize recent data in their calibration of the correlation.
This prioritization of recent data is supported, first, by behavioral findings stating that
 individuals tend to overweight recent information and underweight prior data, second, by
 practitioners, who calibrate their trading strategies with recent data.
 The implementation of this prioritization is achieved by a standard auto-regressive update:
 \begin{eqnarray}
u(t)&=&\alpha\, u(t-1)+(1-\alpha)\, n(t-1)\,\frac{r(t)}{\sigma_r}
\label{eq:u-update}\\
k_{ij}(t)&=&\alpha\, k_{ij}(t-1)+(1-\alpha)\, E_i[s_j(t-1)]\,\frac{r(t)}{\sigma_r}
\label{eq:k-update}
\end{eqnarray}
Choosing $0<\alpha <1$ and with $0< \sigma_r$\footnote
{
$\sigma_r$ is in fact $\sigma_r(t)$, with $\sigma_r(t)^2=\alpha\cdot \sigma_r(t-1)^2 + (1-\alpha)\cdot
\left(r(t-1)-\langle r(t)\rangle\right)^2$ and $\langle r(t)\rangle =\alpha\cdot \langle
r(t-1)\rangle+ (1-\alpha)\cdot r(t-1)$.
}
,  the correct prediction of the sign of the realized stock return $r(t)$
 from a given information source tends to reinforce the trust in that source of information, all
 the more so, the larger the return (scaled by its volatility $\sigma_r$) and the larger the strength
 of the signal. The time scale $1/|\ln(\alpha)|$ sets the memory  duration over which past
 performance continues to impact the adaptive trust coefficients $u(t)$ and $k_{ij}(t)$.
 The update of $u$ and $k_{ij}$ via eq. (\ref{eq:u-update}) and (\ref{eq:k-update}) is performed
 at every time step.

\section{Results of the model}
\label{sec:results}

\subsection{General properties}
\label{sec:generalprop}

Our model is an idealized ``test tube'' representation of a financial market and
 given the simplifications put into the model, we do not aim at reproducing
 faithful statistical characteristics of realistic price dynamics.
 Our objective is to obtain an understanding of how the interplay of
 news, herding and private information can lead to the formation
 of bubbles and crashes. We first point out a few
 properties of the model, that derive straightforwardly from our set-up.

Because we model a closed system, with no new influx of money or stocks after the
 initial endowment of $cash(0)$ and $stocks(0)$, there cannot be any money/wealth creation
 in the long run\footnote{Strictly speaking, the model suffers however from a slight
 money destruction due to the price setting mechanism with the market maker in which the
 log(price) change is linear in the excess demand.  But the number of
 purchased stocks depends on the real price (=exp(log(price))). Therefore, a rapid increase followed by
 a slow decrease of the price decreases the total wealth of the system, by the concavity
 of the logarithmic function. This effect is essentially negligible.}.
 As a consequence, the price trajectory has an upper and lower bound\footnote{The
 upper bound is reached once agents have exhausted all their
 cash. The lower bound is then the agents are all in cash.}.
The constraints on cash and stocks tend to push the price back to its initial value,
 $\textrm{price}(0)=1$, such that the price performs a mean-reverting random
 walk\footnote{The increments of the walk are however not distributed according to a normal law,
 but to a distribution with fatter tails (c.f.
 figure \ref{fig:res}) due to the adaptive strategies of the agents.} around its initial value,
 which will be refereed to in the following as the equilibrium price.

The adaptive process of our agents essentially
 consists in looking for persistent sources of information, which impact on the returns.
 In more detail: for a trade to be profitable, an agent has to first acquire a number of
 asset (at time $t$), then its value has then to increase in
 the following time step, explaining the offset of one
 time step between the information source and the realized
 return in eq. (\ref{eq:u-update}) and (\ref{eq:k-update}).
 The return $r(t+1)$ is however influenced by the information sources at time $t+1$, and not by
 those at time $t$, on which agents based their prediction of $r(t+1)$. This means that, for an information
 source to have some real predicting power, it must have some
 persistence (c.f. Appendix for a more detailed explanation of this mechanism).

\subsection{First results}
\label{sec:first_results}

\begin{figure}[ht]
	\centering
		\includegraphics[width=0.99\textwidth]{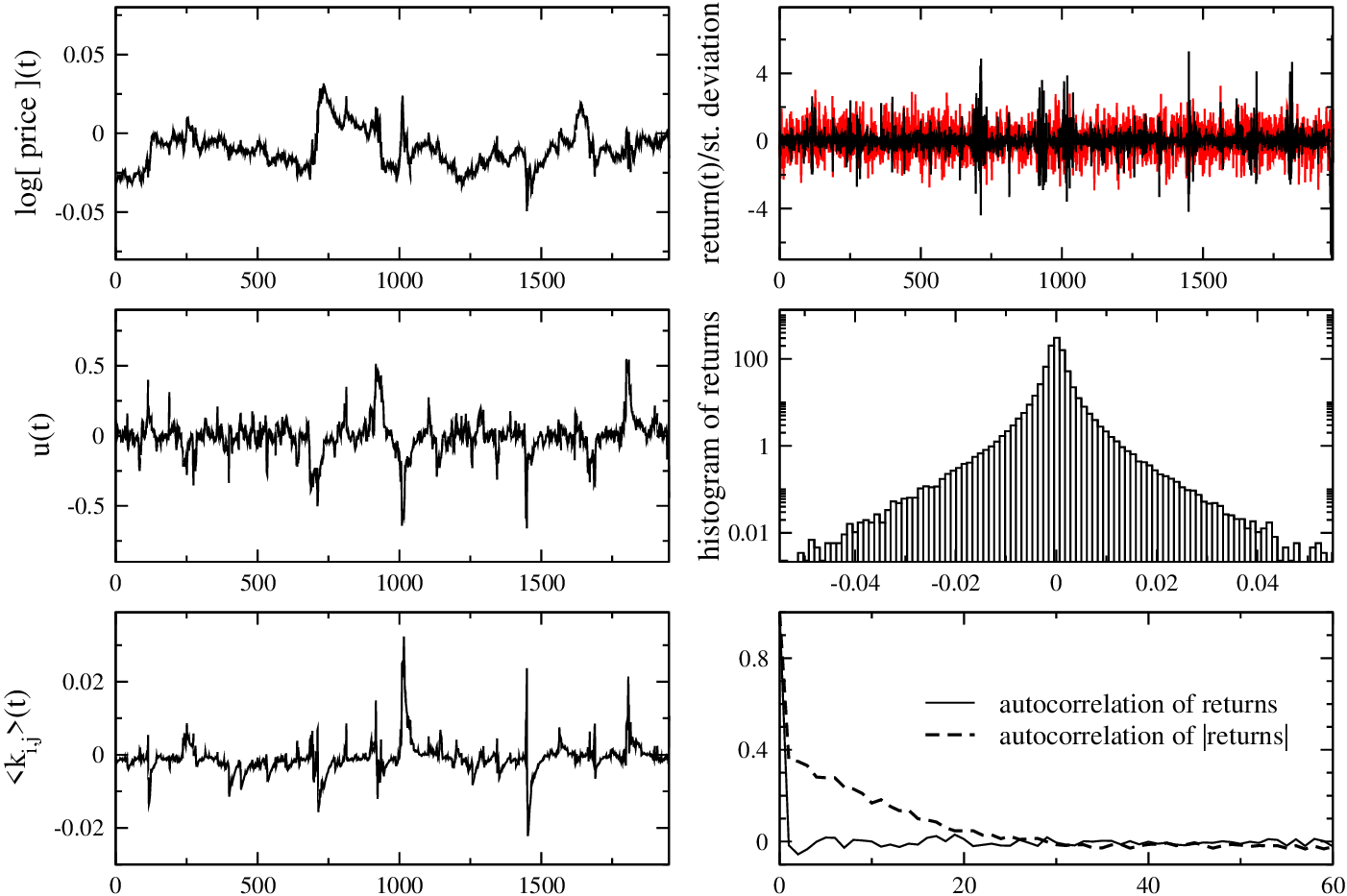}
			\caption{This figure shows a typical realization of the major observables of the system.
 These observables are the time evolution of the price $p(t)$ (upper left panel),
 the one-time-step return $r(t)$ in black with clear evidence of clustered volatility (upper right panel)
 together with the news, $n(t)$, in the background in red,
the news weight factor $u(t)$ (middle left panel)
 and propensity $\langle k_{ij}(t)\rangle$ to imitate (lower left panel). The middle right panel shows the distribution
 of returns: the linear-log scales would qualify a Gaussian distribution as an inverted parabola,
 a double-exponential as a double tent made of two straight lines; in contrast, one can observe
 a strong upward curvature in the tail of this distribution, qualifying a fat-tail property
 compatible with a stretched exponential or power law. The lower right panel shows the absence
 of correlation between returns together with the presence of non-negligible correlation of
 the volatility (here measured as the absolute value of the returns). Note the positive
 value of the correlation of the volatility up to a time about 25 time steps, followed by
 a small negative value up to 80 time steps. The time scale of the correlation of volatility
 is set by the memory factor $\alpha=0.95$ corresponding to a characteristic time scale of
 20 time steps. These results are obtained for $C_1=C_2=C_3=1.0$, and frozen weights
 attributed by the agents to the three information sources drawn out of a uniform distribution
 from 0 to $C_1,C_2,C_3$, respectively. The histogram and the correlation data are computed
 out of a realization with $6\cdot10^4$ time steps.}
				\label{fig:res}
\end{figure}

In our simulations, we fix the number of agents in the system to $N=2500$, the market depth to
 $\lambda=0.25$, the maximal individual conviction threshold to $\underline{\Omega}=2.0$, the
 fraction of their cash or stocks that investors trade per action to $g=2\%$, the initial amount of
 cash and stocks held by each agent to $cash_i(0)=1$ and $stocks_i(0)=1$, and the memory discount
 factor to $\alpha=0.95$, corresponding to a characteristic time of $1/|\ln(\alpha)| \approx 20$ time
 steps. The news are modeled by i.i.d. Gaussian noise. Setting $C_1=C_2=C_3=1.0$,
 figure \ref{fig:res} shows a typical realization of the time evolution of
 the log-price $p(t)$, the one-time-step return $r(t)$, the prediction
 performance of the news, $u(t)$ and the ensemble average of the prediction
 performance of the neighbors, $\langle k_{ij}\rangle(t)$. The middle right panel
 shows the distribution of returns with clear evidence of a non-Gaussian fat
 tail structure. The lower right panel shows the absence of correlation
 between returns together with the presence of non-negligible correlation of the volatility
 (here measured as the absolute value of the returns), which confirms the clear evidence
 of clustered volatility in the time series of one-time-step returns.

While the perceived predicting power of the news, $u(t)$, fluctuates around its
 mean value of $0$, it should be noted that it exhibits significant non-zero values,
 indicating that agents sometimes give a lot of importance to the news.
 If the agents were fully aware of the i.i.d. properties of the news, they would not use
 them\footnote{Recall that the
 return $r(t+1)$ is influenced by the news at time $t+1$ on which agents based their prediction of $r(t+1)$,
 and not by those at time $t$.
 Because the news have no true persistence, they can not have true predictive power.}.
 But because of the adaptive nature of their strategy to the current market regime, agents
 do not use the complete price and news time series to update their trust into the news,
 but only recent data points\footnote{The weight of a data point in the update of $u(t)$ decreases
 exponentially with increasing age with a time scale $\sim 1/|\ln(\alpha)|$.}.
 Due to the use of a finite data set, the i.i.d. news may occasionally show
 persistence\footnote{Our agents do not have a PhD in
 Econometrics and they do not perform proper statistical tests of their hypotheses.},
 leeding to an increase of $u(t)$ as consequence. The statistical fluctuations associated
 with the random patterns that are always presents in genuine noise is misinterpreted
 by the agents as genuine predictability.
 It is the local optimization, that makes the agents see causality,
 where there is only randomness \citep{Taleb2008Fooled}.

The lower left panel of figure \ref{fig:res} shows the average propensity
 to imitate, which also fluctuates around $0$. But, the amplitude
 of these fluctuations is much reduced compared to those of $u(t)$. This is
 because each agent updates individually her propensity to imitate her neighbors
 according to (\ref{eq:k-update}), so that the statistical average
 $\langle k_{ij}\rangle(t)$ is performed over the whole
 heterogenous population of agents, compared with no average for $u(t)$ which
 is common knowledge to all agents.

The crucial parameters of our model are the parameters $C_1, C_2, C_3$, which control the level of
heterogeneity and the a priori preference for the three different types of information.
Changing these parameters changes the way the agents behave in ways that we now explore
systematically.

\subsection{$C_1$-dependence}

Each agent is endowed with a fixed individual preference level, $c_{1i}$, controlling how much
she takes into account the information stemming from the actions of their neighbors.
This level is different from agent to agent, and is drawn from a uniform distribution in the
interval $[0, C_1]$. Thus, the parameter $C_1$ sets the maximal and mean ( $=C_1/2$) innate weight, that agents give
to their social influences.

\begin{figure}[ht]
	\centering
		\includegraphics[width=0.99\textwidth]{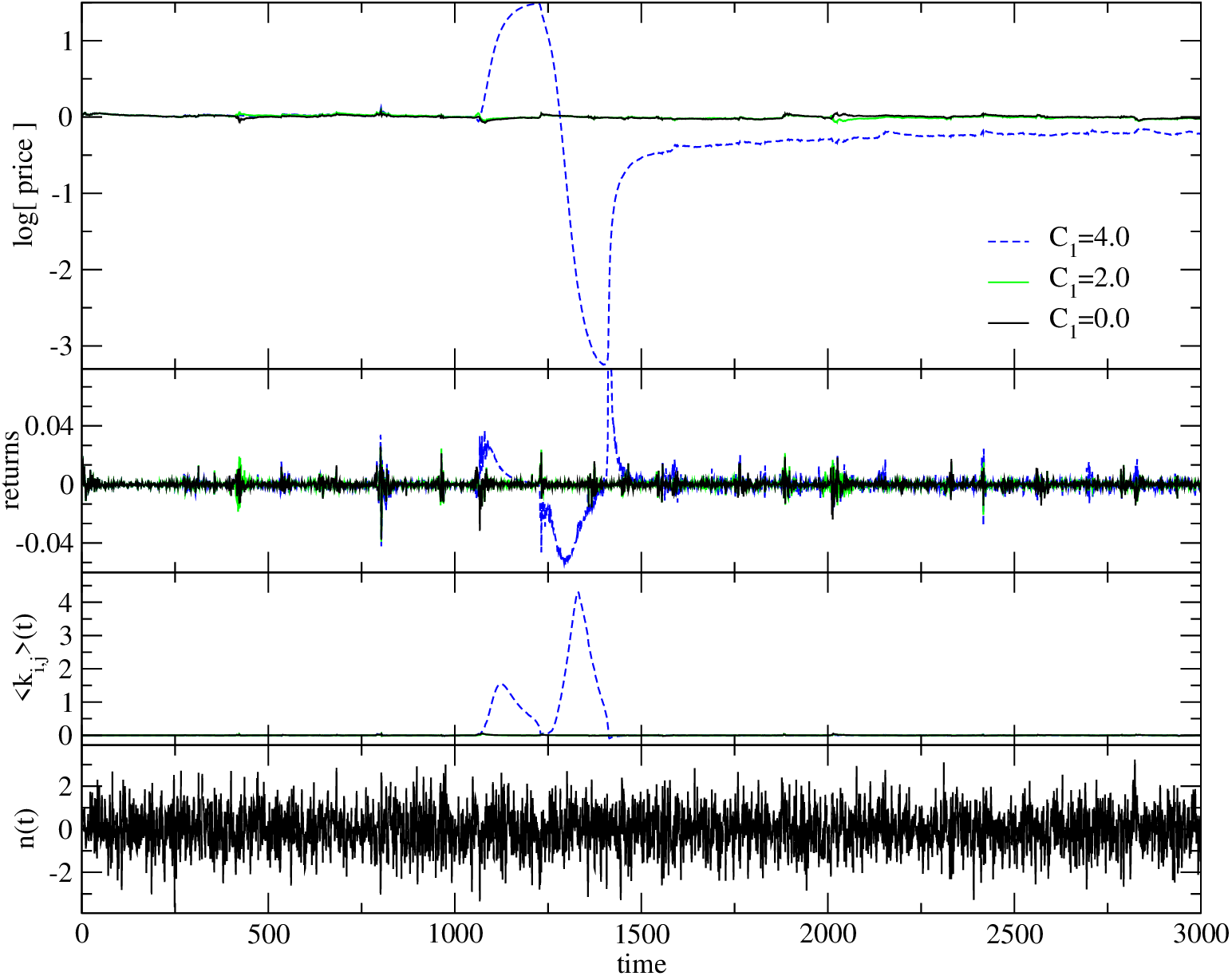} 
			\caption{Evolution of several variables for $C_1=0.0,2.0,4.0$ with the
 other parameters unchanged, including the random seed for the three realizations,
 resulting in the same realization of the news for the different runs (shown in the bottom panel).
 Other parameters are: $N=2500$, $C_2=C_3=1.0$, $\underline{\Omega}=2.0$, $\alpha=0.95$, $\lambda=0.25$, $g=0.02$. }
				\label{fig:price-evol}
\end{figure}

Figure \ref{fig:price-evol}, plots the evolution of several variables
 for three different values of $C_1$, all
 other parameters, including the seed of the random number generator, remaining the same. For vanishing
 propensity to imitate ($C_1=0$), some price spikes can be observed, which are generated by the news
 only, whose influence can be amplified by the positive feedback resulting from adaptation that tends
 to increase the relevance that investors attribute to news after a lucky run of news of the same
 signs. For $C_1=2.0$, one can observe that these peaks are amplified due to the imitation now also
 contributing to the agents' actions. For $C_1=4.0$, a qualitatively different price evolution
 appears. For such large values of the maximal susceptibility to their social environment,
 the price is driven to its
 extremes, its dynamics being only slowed down by the agents' finite cash and stock portfolio
 reaching their boundaries. We show below that this extreme behaviors results from a
 self-fulfilling prophecy, enabled through social interactions.

To better illustrate the effect of increasing $C_1$, the third panel in figure \ref{fig:price-evol}
 shows the average weight factor $\langle k_{ij}\rangle(t)$\footnote{
 $\langle k_{ij}\rangle(t) = \frac{1}{N\cdot J}\sum_{i=1}^N\sum_{j=1}^Jk_{ij}(t)$}
 used by the agents
 to assess the relevance of the information stemming from their neighbors. By increasing $C_1$,
 agents are by definition more susceptible to their neighbors' opinions, making them more likely to act
 in the same way if they show some predictive power. Consequently, since the
 price dynamics is governed by the aggregate demand, herding in opinions leads
 to persistent returns, creating the very returns agents hoped for, which reinforce the prediction power
 of their neighbors in a positive feedback loop. With $\langle k_{ij}\rangle(t)$ and $C_1$ large,
 the opinions of the agents are
 completely shaped by their social component, while the
 news and their idiosyncratic term are essentially ignored.
 Due to this positive feedback loop, a small predictive success of some
 agents can trigger an avalanche of self-fulfilling prophecies, leading to price dynamics completely
 unrelated to the news and to large price deviations from the assets to its equilibrium value.

\subsection{The existence of two regimes}

\begin{figure}[ht]
	\centering
		\includegraphics[width=0.47\textwidth,clip]{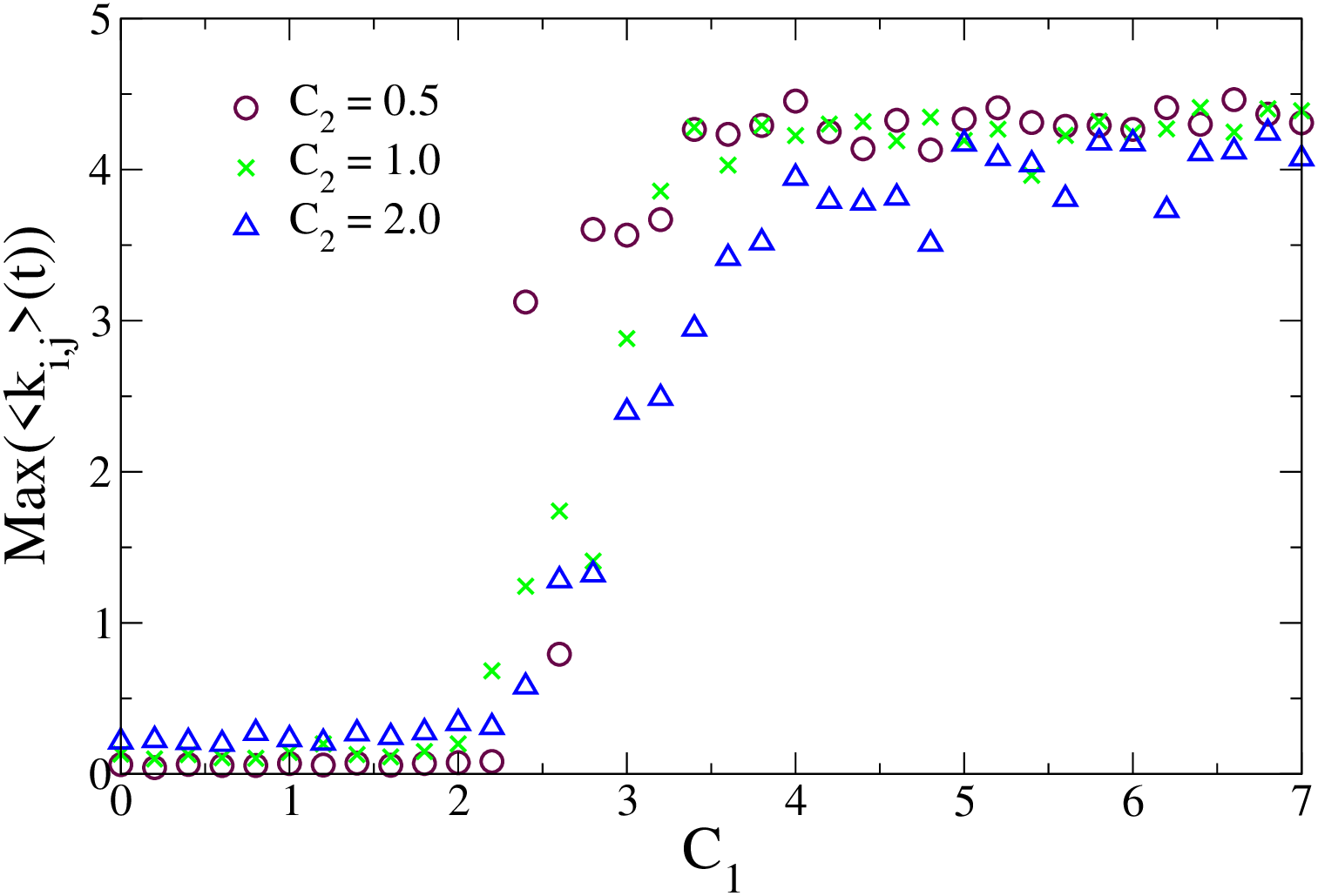}
		\includegraphics[width=0.47\textwidth,clip]{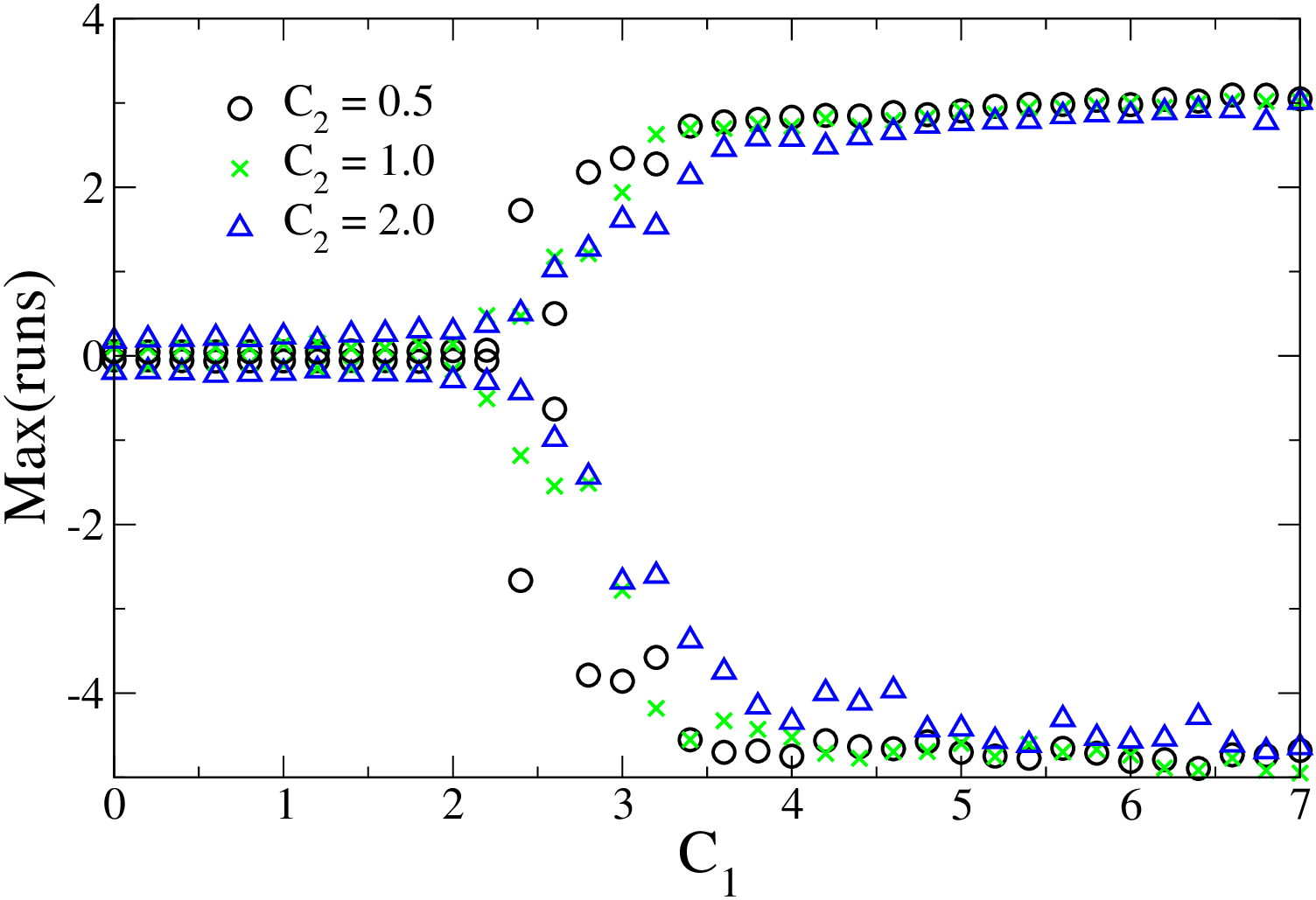}
			\caption{Impact of $C_2$, the innate susceptibility to the news, onto
 the transition from the efficient to the excitable regime in function of $C_1$, the innate
 susceptibility to neighbors' actions. The transitionis measured by the maximal value of
 $\langle k_{ij}\rangle(t)$ (the recent prediction performance of agents' neighbors) (left)
 and the extremal draw-down and -ups (sum of consecutive same signed returns)(right),
 both averaged over many realizations. Both plots show that with increasing $C_2$,
 the critical $C_1$-value increases and the transition smoothens. Other parameters are:
 $N=10^4$, $C_3=1.0$, $\underline{\Omega}=2.0$, $\alpha=0.95$, $\lambda=0.25$, $g=0.02$.}
				\label{fig:C_2-dependence}
\end{figure}

The existence of a bifurcation beyond which a new regime appears is documented in figure
\ref{fig:C_2-dependence}(left), where the maximum value of $\langle k_{ij}\rangle(t)$, averaged over
 many realizations and simulated with the same parameters $C_2=C_3=1$, is plotted as a function
 of $C_1$. One can observe a rather abrupt transition occurring at around $C_1=3$.
 This transition is related to the phase transition of the Ising model, on which our model
 is based. Due to the presence of the adaption induced feedback loops, and with the dynamical character
 of $\langle k_{ij}\rangle(t)$, the precise nature of this transition can not be asserted.
 The existence of this change of regime explains the radical difference of properties
 shown in figure \ref{fig:price-evol} for $C_1=0, 2$ to $C_1=4$. The jump in
 $\langle k_{ij}\rangle(t)$ at $C_1 \approx 3$ is mirrored by a similar transition in the values of
 the maximal draw-downs (sum of consecutive negative returns) and draw-ups (sum of consecutive positive returns)
 as a function of $C_1$ in figure \ref{fig:C_2-dependence}(right).
 For $C_1>3$, a second regime is revealed where very large price moves occur.
 These market events are fundamentally different from the price fluctuations in the
 regime for $C_1<3$. These
 large price changes are reminiscent of the ``outliers'' documented by
 \cite{Johansen1998Stock,johansen_2001} and \cite{sornette_2003}, and recently extended
 into the concept of ``dragon-kings'' \citep{SornetteDragon09}.

\subsubsection{The efficient regime}

For $C_1<3$, agents do not attribute sufficient importance to their neighbors in order to
 trigger the feedback loop that would lead to strong synchronized actions as occurs for large $C_1$'s.
 For small $C_1$, the market is approximately efficient, in the sense that there is no
 autocorrelation of returns, as shown in figure \ref{fig:res}, and the price
 fluctuates rather closely around its equilibrium value. While the major source of fluctuations
 are the news modeled as a Gaussian white noise process, the price fluctuations
 develop strong non-Gaussian features, as a result of the combined effect of
 the adaptive process that tends to amplify runs of same signed news and of the
propensity to imitate that leads to small but non-negligible collective behaviors.

A first interesting conclusion can be drawn that
our model provides a natural setting for rationalizing the excess volatility puzzle \citep{Shiller1981Do},
through the adaptive process of our agents. It could be argued that
our setting is too simplified and unrealistic. But, how do real investors, traders, fund managers
access the value their investment decisions? Necessarily by performing some
kind of comparisons between the realized performance and some benchmarks, which
can be a market portfolio, the results of competitors, the ex-ante expectations, all the above
or others. The adaptive process used by our agents is arguably a simple and straightforward
embodiment of the tendency for investors to adjust their strategies on the basis
of past recent performance, here on how well the news predicted the market returns.
Because measurements are noisy, the resulting estimation leads
unavoidably to an amplification of the intrinsic variability of the news into a much
strong variability of the prices, i.e., to the excess volatility effect.
Somewhat paradoxically, it is the attempt of industrious investors to
continuously adapt to the current market situation, which leads to the dramatic
 amplification of the price volatility. This may be thought of as another embodiment of the ``illusion
 of control'' effect, found in the Minority and the Parrondo games \citep{ satinover_2007_a,
 satinover_2007_b, satinover_2008}, according to which sophisticated strategies are found to
 under-perform simple ones.

\subsubsection{The excitable regime}

A population of agents characterized by $C_1>3$, represents the situation in which many
 agents know that their idiosyncratic information and the
 news are incomplete. In order to compensate for this lack of information, agents tend
 to imitate the actions of successful acquaintances. Under these
 conditions, the average propensity to imitate, $\langle k_{ij}\rangle(t)$, exhibits
 extreme values, resulting in large price deviations from the equilibrium price and
 periods of persistent returns, as shown in figure \ref{fig:price-evol}. In this regime, the
 market is in an excitable state. By imitating the opinions of recent winners who profited
 from some departure of the market price from its equilibrium value,
 our agents tend to amplify this anomaly, further strengthening the attraction of this
 strategy for other agents, eventually ending in a bubble and crash.

 The triggering event responsible for the increasing weight that agents entrust to their neighbors' opinions
 is nothing but the  random occurrence of a sequence of same signed news.
 As explained in section \ref{sec:first_results},
the weight $u(t)$ of the news in their opinions is increased when the agents perceive a
pattern of persistence in the news, which also induces persistent returns.
 Then, the agents reassess their belief and give more
 importance to the news.
Because the pattern of persistence of the news is common knowledge, this tends
to align the decisions and actions of the agents. As a consequence, their aggregate
impact makes happen the very belief that initially led to their actions, thus
increasing the prediction power of the agents' opinions. As a whole, the agents
see that the opinion of their friends is accurate, thus tending to increase their trust.
 This increase in the propensity  $\langle k_{ij}\rangle(t)$ to imitate can
lead to a cascade of trading activity, resulting in the rise of a bubble, as described in
details in the appendix. This scenario is a cartoon representation of the
well-documented fact that
 many bubbles start initially with a change of economic fundamentals. Translated
 in our agent-based model, this change of economic fundamentals is nothing but
 the streak of same signed news that tells the story of an increasing market (for positive news). This
 small positive bias can be sufficient to nucleate a process that eventually
 blossom into a full-fledged bubble. The corresponding amplification of the news
 put the price on an unsustainable trajectory. This occurs especially when the system lives
 in the excitable state, in which the price can easily overshoot the
 values implied by the good/bad news.

Once such a cascade has begun and the best strategy is to follow the herd, agents are, in the case
 of a bubble, buying stocks at every time step and pushing the price up till they have no money
 left to further increase the price. At this point, their predicting power decreases due to they
 decreasing impact on the returns and the cascade ends. As financial bubbles feed on
 new money pouring in the market, the lack of new liquidity is a well-known
 factor of instability for financial bubbles \citep{Kindleberger2005Manias,VernonSmith08}.
 Following this buying phase, the portfolio of agents
consists mainly of stocks, biasing their actions towards selling.
 Now, some randomly occurring negative news are sufficient
 to trigger a reverse cascade, the crash, leading to an overshoot of the price
 below its equilibrium value.
 This scenario provides a clear distinction between the fundamental cause of the crash
 (the unstable high position of the price that has dried up all liquidity available)
 and the triggering proximal factor (a random occurrence of a sequence of negative news).
 In line with many observations, crashes in our model do not need a dramatic piece
 of negative information. Only a trickle can trigger a flood once the market as a whole
 has evolved into an unsustainable unstable position.

Due to the symmetry between buying and selling in our model, the price can, starting from
 its equilibrium value, depart in either direction, creating either a bubble (over-valuation of
 the asset over an extended period of time) or an negative bubble (under-valuation of
 the asset). This deviation will then be ended by either a crash (fast drop in price after
 a bubble) or a rally (fast appreciation after a negative bubble). In our analysis, we concentrate on the
 case of bubbles followed by a crash, because this is the more common scenario. The reason
 for this is twofold. First, in real markets, short-selling can occur, but is not equally
 available to all market players.  The second reason has behavioral origins. In a bullish
 regime, people are progressively attracted to invest in financial markets, tending to push the price upward.
Once invested, their attention is more focused on the financial markets. Fear and greed
often lead to over-reactions and possible panics when the sentiments become negative,
triggering herd selling which self-fulfills the very fears at their origin \citep{Veldkamp2005Slow}.

\begin{figure}
	\centering
		\includegraphics[width=0.8\textwidth]{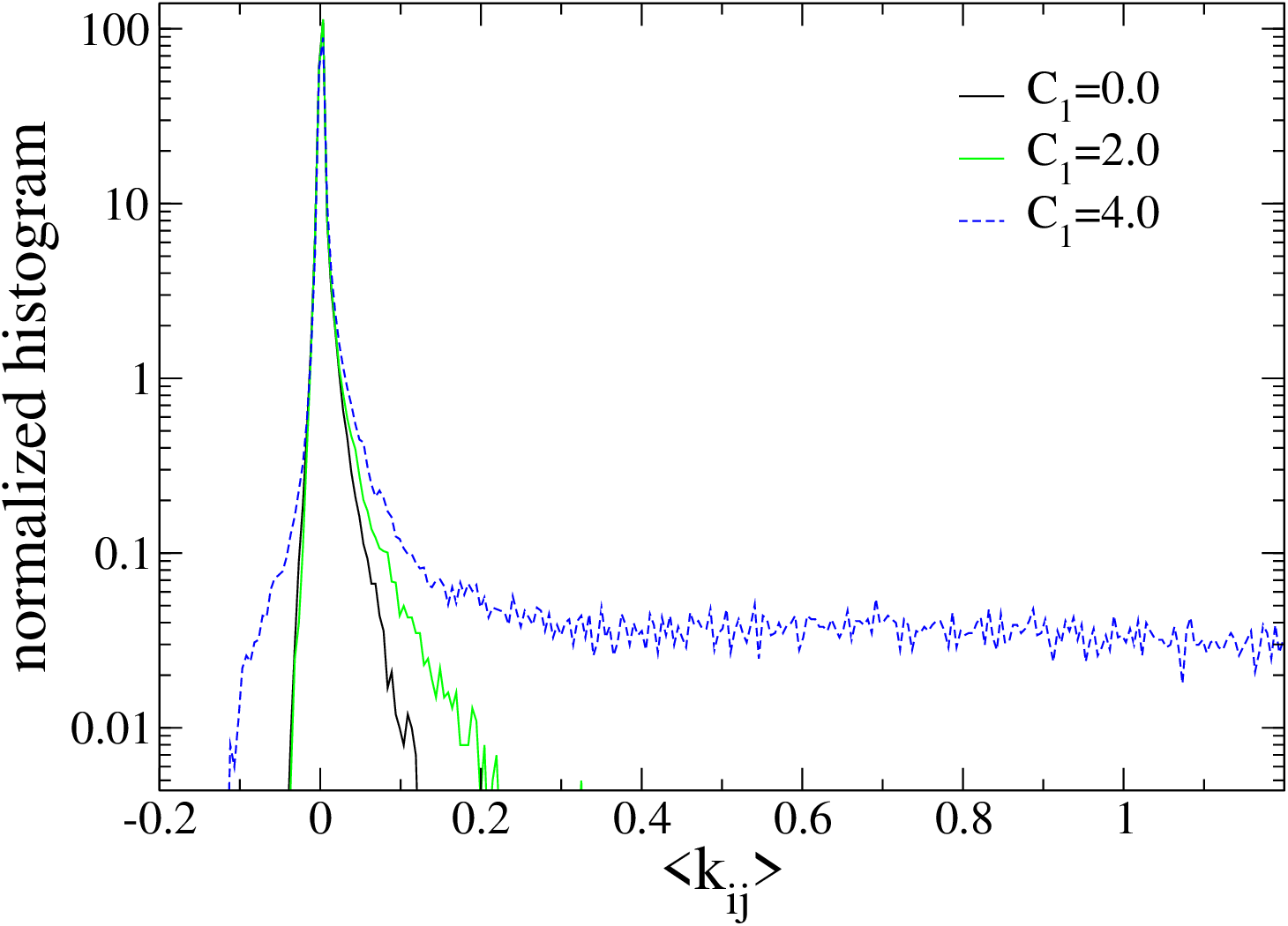}
			\caption{Normalized histogram of $\langle k_{ij}\rangle(t)$
 for different values of $C_1$.}
				\label{fig:av_k-histogram}
\end{figure}

Another interesting characteristic of the herding regime occurring for $C_1>3$ is that it is very difficult to
diagnose this regime from the properties of the price recorded outside those transient episodes
of booms or crashes. Indeed, outside these special moments of ``exuberance'', the market behaves
as if in the regime $C_1<3$. Bubbles and crashes do not belong to the normal regular dynamics
of the model. They are only experienced when certain conditions are fulfilled, as explained above, that combine
to create these transient instabilities. They can thus be considered as ``outliers''
in the sense of \cite{Johansen1998Stock,johansen_2001}, or using a better
more colorful terminology, they are ``dragon-kings''
\citep{SornetteDragon09}. The statistical analysis of the distribution of f $\langle k_{ij}\rangle(t)$
confirms this claim. Figure \ref{fig:av_k-histogram} shows the appearance of an extremely fat tail in the
distribution of $\langle k_{ij}\rangle(t)$ over the ensemble of different realizations as a
function of time for $C_1=4$, while its bulk remains approximately identical to the distribution
obtained for the smaller values of $C_1$ below the critical threshold $\simeq 3$. This confirms the
existence of a class of transient regimes, the booms and crashes, which coexisting
with the normal dynamics of the prices.

The hidden nature of the regime associated with $C_1>3$ and the random occurrence of
 triggering news lead to the prediction that advanced diagnostics of bubbles and
 crashes should lead to numerous false alarms. Consider the study of
 \cite{Kaminsky1998Currency}, who has compiled a large list of
 indicators of financial crises, suggested
 by the fundamentalist literature on the period from 1970 to 1995 for 20 countries.
 Out of the 102 financial crises in her database, she finds that the specificity
 of the indicators is quite low: only $39\%$ of the ex-ante diagnostics coincided with a
 crisis, suggesting that fundamental reasons should be expanded by behavioural ones to
 explain the emergence of crises.

We thus come to our second important conclusion: the present model provides a simple mechanism
for the existence of two populations in the distribution of prices, exemplifying
the concept that booms and crashes are qualitatively different from the rest of the price moves.
The second population of boom-crash (dragon-kings) appears when the innate
propensity to herd reaches a threshold above which a self-reinforcing positive feedback loop
starts to operate intermittently.

\subsection{$C_2$-dependence}

Figures \ref{fig:C_2-dependence} displays the impact of $C_2$, the a priori importance of the
 news, onto the transition from the efficient to the excitable regime. Both panels show
that the more the agents trust the news, the stronger $C_1$ has to be for the system
 to become excitable. With increasing $C_2$-values, we also observe that the transition
 becomes smoother and the maximum $\langle k_{ij}\rangle(t)$ is decreased.
 This corresponds well to the intuition that, if traders
 are well informed and believe that the news correctly describe the economy, such drastic
 over- and under-valuations are less likely to happen and a higher level of
 panic is needed for a crash to happen.

\subsection{$\alpha$-dependence}

Although the presence of the $\alpha$-parameter
is crucial, its specific value (within a certain range below $1$) has only a minor
 importance.
Recall that $\alpha$ sets the time-scale of the market regime, since
 it controls the length of the time series that is used to estimate the predicting power of
 the different information sources.
The fact that $\alpha<1$, i.e. that agents' strategies
 are designed to identify local market regimes is the reason that makes them
 possible in the first place. It is the local adaptation, which is the true
 origin of a bubble and a subsequent crash.

\begin{figure}
	\centering
		\includegraphics[width=0.47\textwidth,clip]{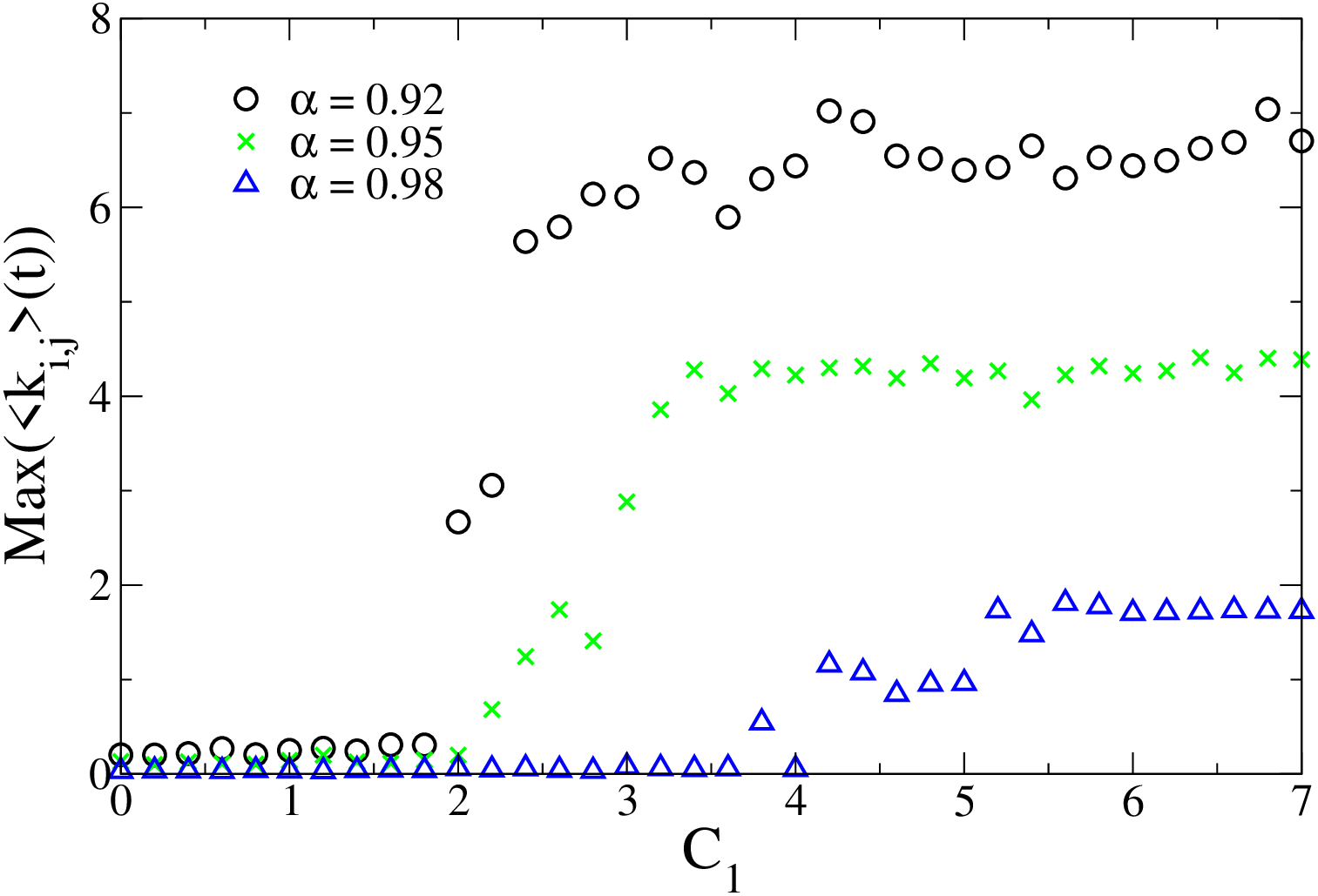}
		\includegraphics[width=0.47\textwidth,clip]{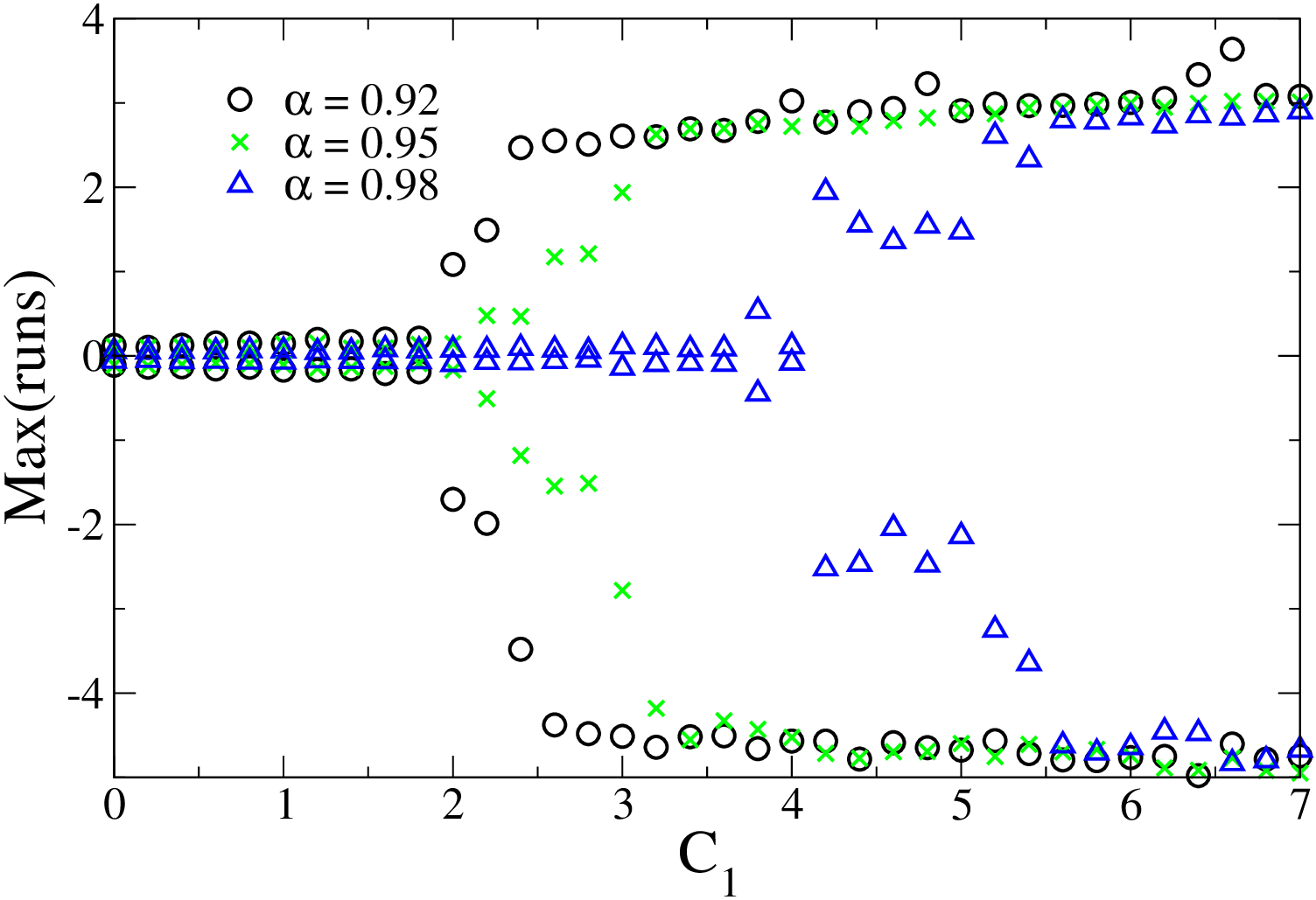}
			\caption{Same as figure \ref{fig:C_2-dependence}, except that $C_2$
 is fixed at 1.0 and that the impact of $\alpha$, which fixes the length of the time-span
 over which agents measure the predicting power of the different sources of information,
 is investigated. For larger $\alpha$, the critical $C_1$-value increases, the transition
 smoothens and the largest possible $\langle k_{ij}\rangle(t)$ values is decreased.Other
 parameters are: $N=10^4$, $C_3=1.0$, $\underline{\Omega}=2.0$, $\lambda=0.25$, $g=0.02$.}
				\label{fig:alpha-dependence}
\end{figure}

Figure \ref{fig:alpha-dependence} shows the impact of $\alpha$ on the transition
 from the efficient to the excitable regime. The closer $\alpha$ is to $1$, the larger
 the critical $C_1$ for which the system becomes excitable. With a larger memory, the
 growth of the propensity
 $\langle k_{ij}\rangle(t)$ to imitate is more limited  because the agents see now much
 better the bigger picture and are less easily carried away by a temporal
 coordination of their neighbors. Changing $\alpha$ leaves
the maximal drawn-downs and -ups unchanged because, once an coordination of the agents
 starts, the only way to stop it is via the drying up of their cash/stock-reservoir.

\subsection{Alternative clearing condition}

We have played with variations of the implementations of the clearing conditions with different
market maker' strategies. For instance, when the market make is adapting the price
after (rather than before) the exchange of assets with the agents, we find the same
qualitative results and bubbles and crashes occur by the same mechanism.
 With such an price clearing condition however, news bear real
 predictive power, destroying the efficiency of the market.

In general, the present model is very robust with respect to changes in the different
 ingredients. As long as agents can interact and are locally optimizing their
 strategies, bubbles and crashes do appear.

\section{Conclusion}
\label{sec:concl}

In this paper, we have addressed two major questions:
\begin{list}{-}{ }
\item Why do bubbles and crashes exist?
\item How to they emerge?
\end{list}

We approached these questions by constructing a model of bounded rational, locally optimizing
 agents, trading a single asset with a very parsimonious strategy. The actions of the agents
 are determined by their anticipation of the future price changes, which is based on
 three different sources of information: private information, public
 information (news) and information from their neighbors in their network of professional
 acquaintance. Given these information, they try to maximize their usefulness by constantly
 scanning the market and adjusting the weight of the different sources to
 their opinion by the recent predicting performance of these sources. In this way, they are
 always adapting
 their strategy to the current market regime, such that they can profit from an opportunity if it arises.

We find that two regimes appear, depending on how strong the agents are influenced by their
 neighbors (controlled by the parameter $C_1$). In the regime of small $C_1$'s, the low
 herding/efficient regime, agents are sometimes more influenced by the news and sometimes
 more by their neighbors, but due to the small level of trust they put into their neighbors
 by default, they do not get carried away in over-imitating their neighbors if the latter,
 for a short time interval, seem to be good predictors. The returns are mostly driven by the
 global and idiosyncratic news. The resulting market is approximately efficient, with
 the price not deviating much from its equilibrium value.

We find that the return distribution is however quite different from that describing
 the exogenous news. Our simple agents are able to transform the string of independent
 normally distributed news (both for the global and idiosyncratic news) into a return
 distribution with fatter than exponential tails, showing a clear sign of excess volatility.
 Also clustered volatility and a non-zero autocorrelation in the volatility of the returns
 are observed
 while the returns themselves remain uncorrelated, in agreement with absence of arbitrage
 opportunities (at least at the linear correlation level). These different properties show
 that our simple model can reproduce some important stylized facts of the stock market, and
 can motivate the possibility to test its prediction in other market regimes.

By increasing $C_1$ above a certain critical value, the system enters a second regime
 where the agents give on average more importance to their neighbors' actions than to
 the other pieces of information. By increasing the awareness of their neighbors' actions,
 agents are more likely to coordinate their actions, which increases the probability that
 the direction of the return results in the predicted direction, which then again increases
 their trust in these successful predicting neighbors. Due to this positive feedback loop,
 the average coefficients $\langle k_{ij} \rangle$ (the dynamic trust of agent $i$ in agent $j$)
 can surpass a critical value and the agents' opinions are dominated by only this information
 term, resulting in series of consistently large same-signed returns. Because the agents are
 always trying to maximize their returns, it is rational for the agents to follow the majority
and to ``surf'' the bubble or the crash. This regime is characterized
 by large deviations from the equilibrium price resulting from a coordination of the agents'
 actions due to their local adaptation of their strategy to the mood of the market. Not only
 is it rational to follow the herd, we have also showed (in the appendix) that the agents who
 are early imitators of their successful neighbors in the early
 stage of a bubble/crash are those who will
 accumulate the largest wealth among all the agents after
 the market has returned to its normal regime.

We showed that the origin of these large deviations from the equilibrium price
 nucleate from the news. A random occurrence of a sequence of same signed news pushes the
 price in one direction and starts the coordination process of the agents. This situation
 is reminiscent from the mechanism for the initiation of real world bubbles, where an innovation leads to a period
 with a majority of positive news, which also move the market.
 Because the reason of the positive market move is an innovation, the agents are not entirely
 sure of its intrinsic value and seek advice from
 some of their professional collegues. If those colleagues tell them that they made large
 profits with this asset and they trust these colleagues, they will follow their advice,
 resulting in the same kind of behavior as produced by our model.

By following each other's actions, the agents push the price up, beyond its equilibrium value, up to
 an unsustainable level. Once the hype has cooled of and the agents have invested all their
 cash into the stock, just a little push by negative news can cause the price to collapse,
 resulting in a crash, without any apparent reason.

By increasing the prior propensity $C_1$ to imitate to a high value,
the average behavior and properties of
 the dynamics of the model is unchanged. Outside of these large price variations occurring during
 rare bubbles and crashes, the dynamics looks similar to that documented in the low $C_1$
 regime, i.e., appears to function like an efficient market. Therefore, attempting to estimate
 the value of $C_1$ just from the normal price dynamics is essentially impossible.
 The occurrence of a bubble/crash is an event that has drastically different statistical
 characteristics than the normal price fluctuations, exemplifying the occurrence of
 ``outliers''  (or ``dragon-kings'')
 that have been documented empirically for financial draw-downs \citep{Johansen1998Stock,johansen_2001,sornette_2003}.

\vskip 0.5cm
{\bf Acknowledgments}: We would like to thank Wei-Xing Zhou for invaluable discussions
during the course of the project and Gilles Daniel and Ryan Woodard for a critical reading
of the manuscript.



\clearpage

\clearpage

\clearpage

\clearpage

\clearpage

\appendix
\title{Appendix}

\section{Detailed analysis of the emergence of a bubble}
\label{sec:news-impact}

To clarify the mechanisms leading to the dynamics of the here presented model, we now illustrate
 some details on the micro-scale dynamics of the model. First we will go step-by-step through
 an occurrence of increased volatility, shown in figure \ref{fig:news-impact_C1=1}, with the
 system being in the efficient regime and explain in detail the relationships
 between the different variables.

Second, we will investigate the emergence of a bubble in the excitable regime and compare
 it to the dynamics resulting from the same stream of news in the efficient regime in
 figure \ref{fig:news-impact_C1=1+3} and \ref{fig:news-impact_bigpic}.

\subsection{Step-by-step description of the dynamics in the efficient regime}

\begin{figure}[ht]
	\centering
		\includegraphics[width=0.99\textwidth]{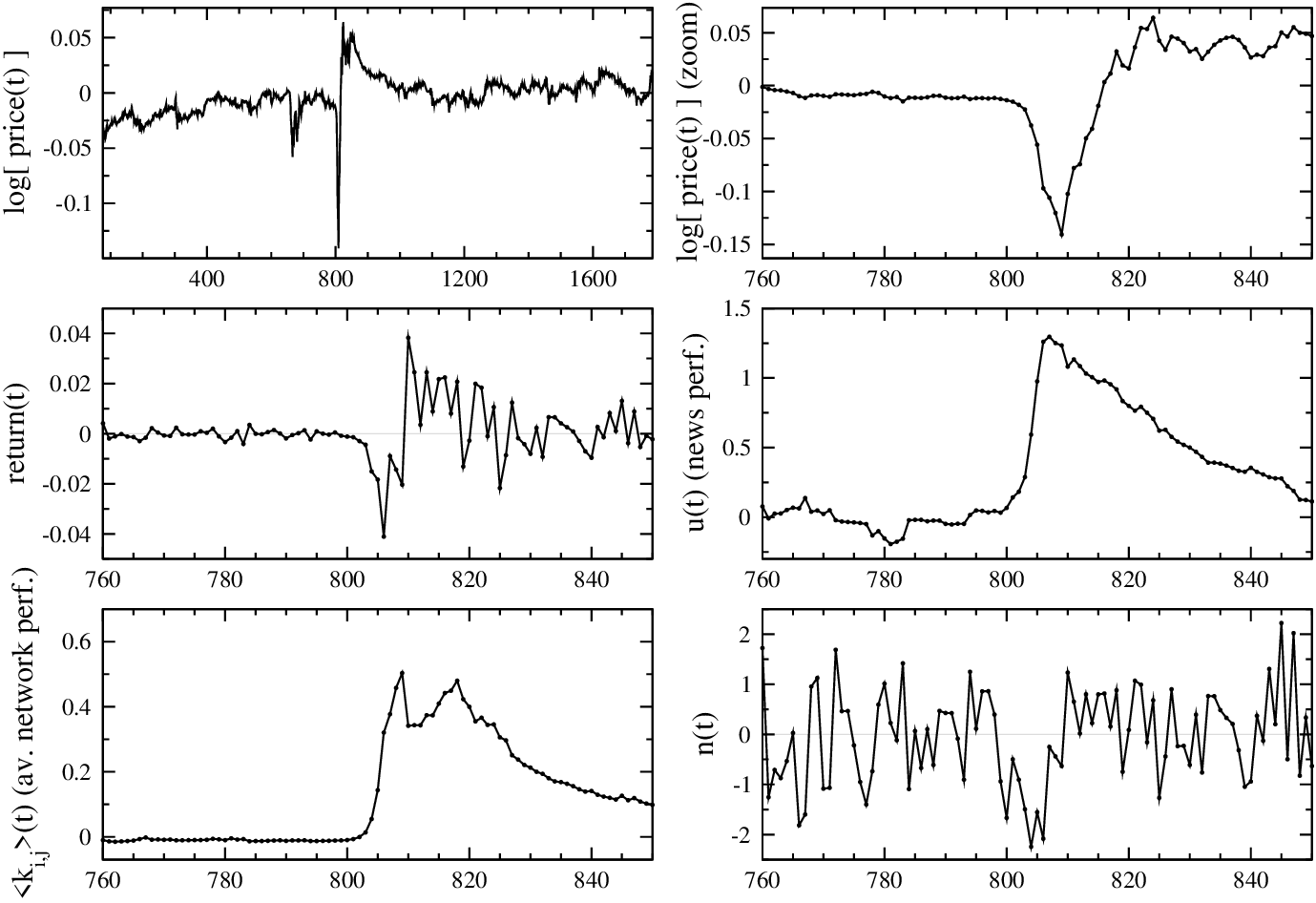}
			\caption{Time series of several key variables showing
 the response to the news for $C_1=1$ (efficient regime) and $C_2=C_3=1.0$. Upper left
 panel: a portion of the price time series with a drop and rebound. Upper
 right panel: a magnification of the upper left panel around the increased volatility. Middle left
 panel: the time series of the returns. Middle right panel: the weight $u(t)$ of the news
 showing a fast growth over the time interval in which the news are all negative, followed
 by a decay over a time scale given by $1/\ln(\alpha) \approx 20$ time steps. Lower left
 panel: The average weight $\langle k_{ij} \rangle(t)$ of the propensity to imitate also
 exhibits a fast acceleration followed by a slower decay. Lower right panel: the time series
 of news, generated as a white noise, which can nevertheless exhibit runs of same-sign values.}
				\label{fig:news-impact_C1=1}
\end{figure}

Figure \ref{fig:news-impact_C1=1} displays the dynamics of the key variables around the time
 $t=800$, where the price suddenly, crashes, rebounds and then slowly relaxes to its
 pre-existing level. An increase of $u(t)$, the news' performance and $\langle k_{ij} \rangle(t)$,
 the average weight used by the agents to assess the relevance of the information stemming from
 their neighbors, is occurring at the same time.

The origin of this burst can be traced back to the random occurrence of a sequence of same
 signed news, shown in the lower right panel of figure \ref{fig:news-impact_C1=1}. Recall
 that we assume that the news are independently and identically distributed. Thus the dip
 structure in the news' realization is purely ``bad luck'', i.e. a stream of small bad news
 impact the market. The response of the agents to these run of bad news develops as follows.
The observation of the news $n(t)$ gives the agents an information about the next return $r(t+1$), but in order to profit from
this insight, the agents have to act before $t+1$, i.e. they use $n(t)$ to buy or sell at time $t$. Therefore, a
burst of activity, which has its origin in the news, can only occur if the sign of the news is, by chance, the same for
several time steps as it is the case from $t=799$ to $809$.

Let us report minutely the micro dynamics of the model to better understand this burst of activity.
 At $t=799$, the news turns out to be negative, which suggests to the agents that the price may
 drop from $t=800$ to $t=801$. To prevent their portfolio from losing in value from time $t=800$
 to $t=801$, some agents reduce their exposure to the market and sell a fraction $g$of their assets at
 $t=799$. If enough agents listen to the news, as it is in this case, this selling will result in
 a negative return from $t=799$ to $t=800$. Then at $t=800$, the news is, by chance, again negative
 resulting again in a negative return from $t=800$ to $t=801$. This negative return confirms the
 negative news from $t=799$, leading agents to increase $u(t)$, the weight they attribute to the news.

The exponential growth of the weight $u$ continues as long as the sequence of negative news goes
 on, further amplifying the impact of the news on the agents' decision and therefore on the price.
 Note that the average weight $\langle k_{ij} \rangle(t)$ of the propensity to imitate also
 exhibits a fast acceleration. This is due to the fact that the agents find that imitation is
 also a good predictor of the returns, since a majority of agents are following the news and
 are trading into the same direction. By this process, there is an amplification of the response
 of the whole herd to the exogenous news. When the run of bad news stops, it takes about
 $\simeq 1/\ln(\alpha) \approx 20$ times steps for $u$ to relax back to its previous value. In
 this example, the maximum of $u(t)$ occurs at $t=807$. At $t=808$, $u$ decreases lightly due to
 the small amplitude of the news at $t=807$. Once $u$ has reached a certain level, the news
 completely dominates agents opinion and thus also determines the returns.

At $t=810$, the sequence of negative news is terminated by positive news, resulting in a large
 positive return due to the large value of $u(t)$. Furthermore, as the news at $t=809$ predicted a
 negative return at $t=810$, the predictive power of the news seems to have decreased, having a
 a decrease of $u(t)$ as consequence. Now that the news resumes its usual random switching signs
 and it is no longer a good predictor of the return, $u$ decreases exponentially.

This case study illustrates that the occurrence of bursts of price variations is nothing but the amplification of runs
of same-sign news, which leads to an exponential growth of the news weighting factor $u$, which itself increases dramatically
the sensitivity of the agents to all future news. This heightened sensitivity lasts over a characteristic scale determined by
the coefficient $\alpha$ governing the memory of the adaptation process (eq.~\ref{eq:k-update}). The process
of agents' adaptation to the news and information from their neighbors, together with the
random lucky or unlucky occurrence of runs of news of the same quality, is at the origin of the occurrence
of this period of increase volatility.

\subsection{Efficient regime vs excitable regime}

\begin{figure}[ht]
	\centering
		\includegraphics[width=0.99\textwidth]{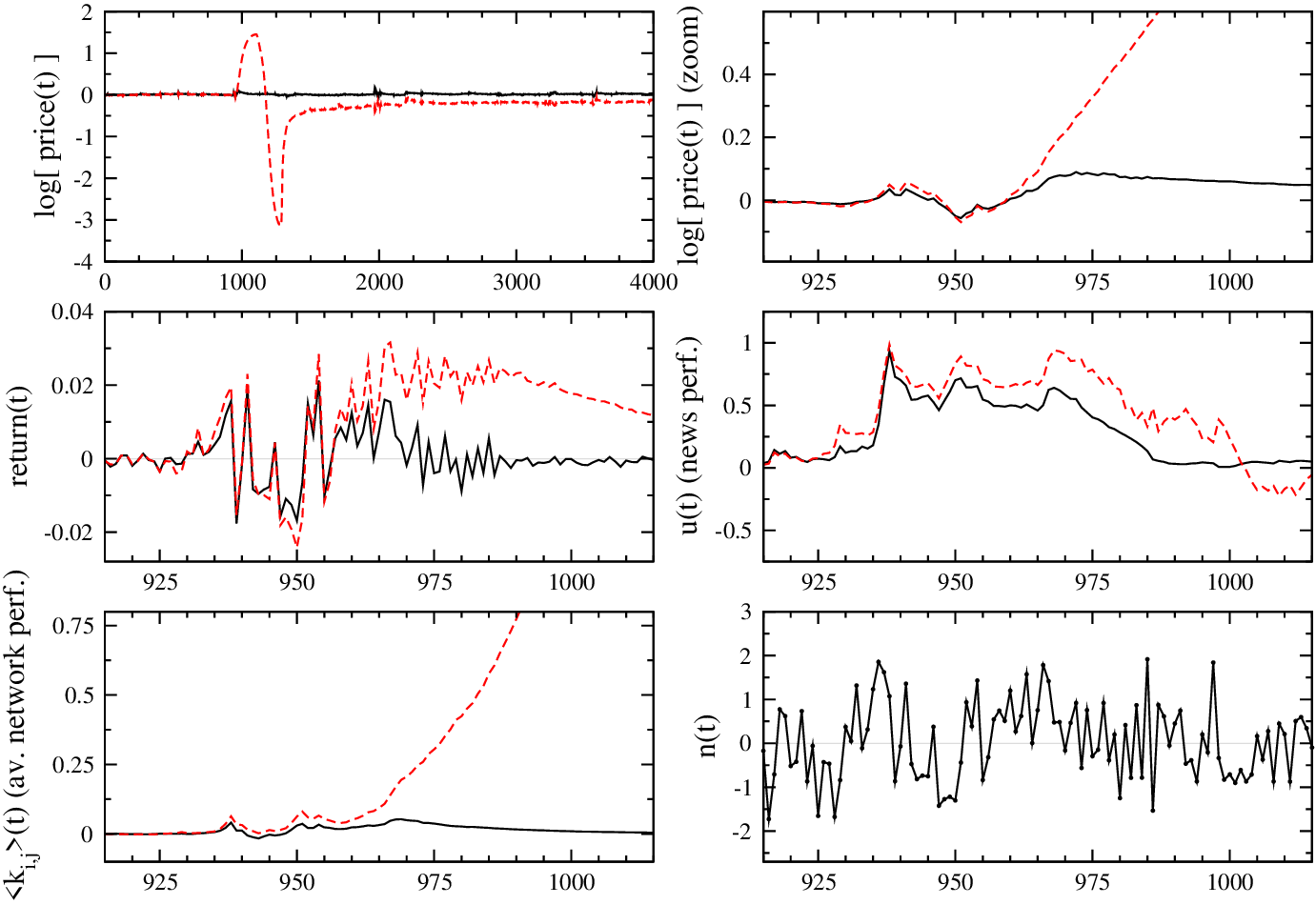}
			\caption{Time series of several key variables showing the
 response to the news for $C_1=C_2=C_3=1.0$ (efficient regime, in black) and $C_1=4$ with
 $C_2=C_3=1.0$ (excitable regime, red dashed). We can observe that the random occurrence of
 persistence in the news stating around $t=930$, initiates a bubble if agents
 give too much weight to their neighbors' actions.}
				\label{fig:news-impact_C1=1+3}
\end{figure}

\begin{figure}[ht]
	\centering
		\includegraphics[width=0.99\textwidth]{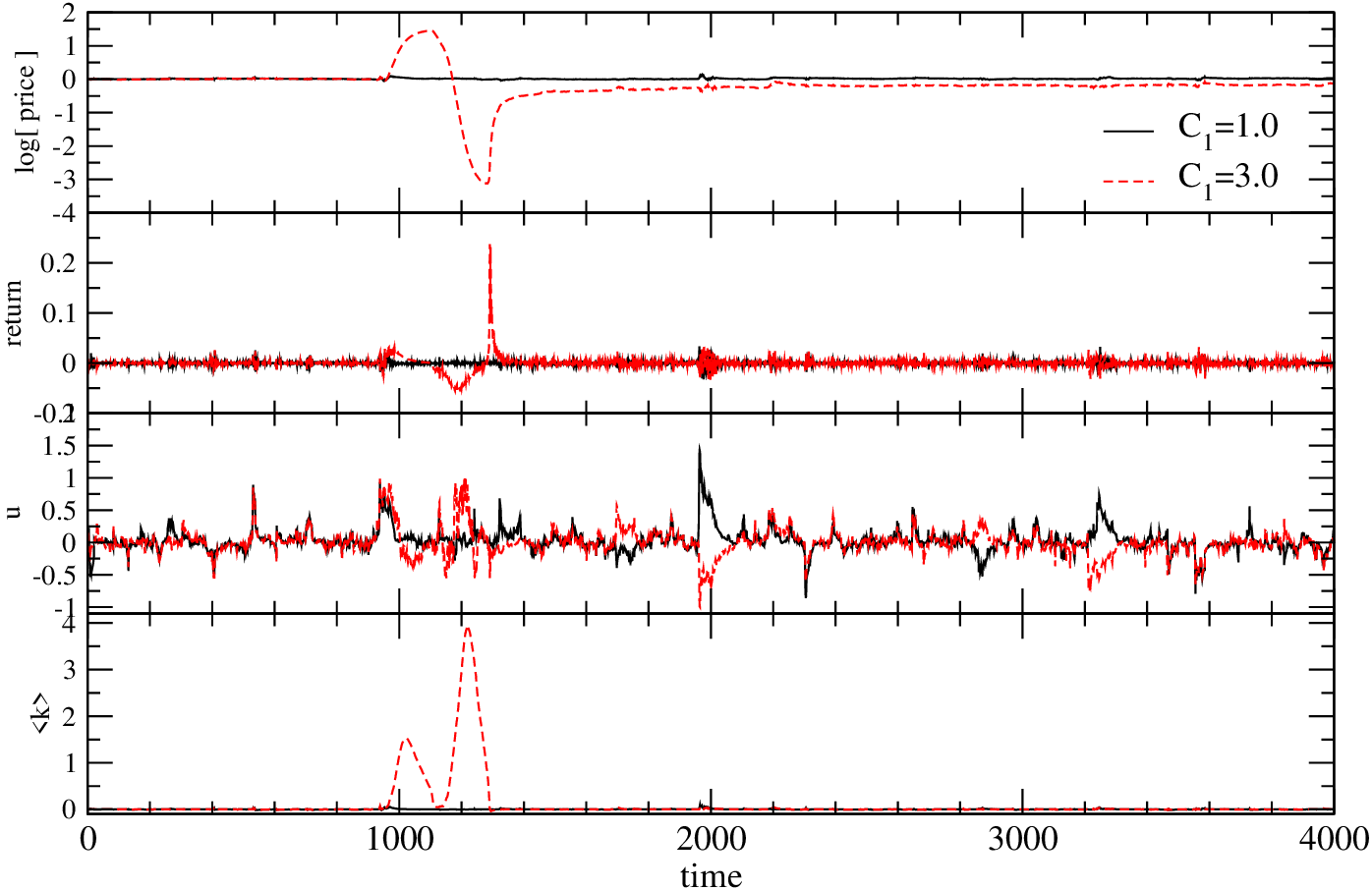}
			\caption{Magnification of the realization of the crash shown in figure
 \ref{fig:news-impact_C1=1+3}. Top to bottom: plots of the price, return, activity, news weight factor
 and average imitation factor, as a function of time.
			}
				\label{fig:news-impact_bigpic}
\end{figure}

In figure \ref{fig:news-impact_C1=1+3}, we plot the detailed dynamics during the nucleation
 of a bubble. The black continuous lines display the evolution of several variables with
 the system being in the efficient regime, i.e.  $C_1=C_2=C_3=1.0$. The red dashed lines
 represent the same variables, with all parameters unchanged (including random seed),
 except that agents susceptibility to their neighbors' actions is increased
 ($C_1=4.0$) such that the system is in the excitable regime.
 Figure \ref{fig:news-impact_C1=1+3} shows the detailed nucleation of the bubble, whereas
 figure \ref{fig:news-impact_bigpic} shows the dynamics on a larger time scale.

In the second panel on the left in figure \ref{fig:news-impact_C1=1+3}, we plot the evolution
 of the return and witness a burst of volatility starting around $t=930$. The origin of this
 volatility can be attributed to a random occurrence of some `persistence' in the news $n(t)$,
 as explained in the previous section. This persistence increases the news' prediction power $u(t)$
 and, because all agents are subject to the same news, agents' actions tend to synchronize, inducing
 an increase of the prediction power of their neighbors, $\langle k_{ij} \rangle(t)$.

Up to
 $t=957$, the dynamics of the system in the efficient regime only deviates marginally from those
 of the excitable regime. After $t=957$, their differences become apparent. In the
 excitable regime, where the neighbors' influence is a stronger factor in the opinion
 formation, the long string of positive news from $t=957$ to $t=969$ is able
 to increase the average interaction weight $\langle k_{ij} \rangle(t)$ up to a level high enough,
 such that the opinion of the agents, and therefore also their actions, are dominated by their
 neighbors'. As a consequence, the price continues to increase even after the positive news sequence has ended. In the efficient
 regime, on the other hand, the volatility of the returns, $u(t)$ and $\langle k_{ij} \rangle(t)$
 return to their normal values after the `luck streak' of positive news.

In the excitable regime, as a consequence of the strong propensity to interact , once
a price rally or a crash is started, the dominating impact of the herd both in
the adaption process and in the price impact makes the price trend
self-reinforcing and basically independent of the sign of the news. This
explains the very large amplitude of the price deviation compared with the price in
 the efficient regime in figure  \ref{fig:news-impact_C1=1+3} and \ref{fig:news-impact_bigpic}.


\newpage

\clearpage

\end{document}